\documentclass[twocolumn,prb,aps,showpacs,floatfix,superscriptaddress]{revtex4}
\usepackage{graphicx}
\usepackage{verbatim}

\begin{document}

  \newcommand{\wert}[3]{$#1=#2\,\rm{#3}$}

\title{Superconductivity-Induced Distortions of Phonon Lineshapes in Niobium}

\author{F. Weber}

\affiliation{Materials Science Division, Argonne National Laboratory, Argonne, IL, 60439, USA}
%\email{frank.weber@kit.edu}

\author{L. Pintschovius}

\affiliation{Karlsruher Institut f\"ur Technologie, Institut f\"ur
Festk\"orperphysik, P.O. Box 3640, D-76021 Karlsruhe, Germany}

\begin{abstract}
Superconductivity-induced changes of phonon lineshapes in niobium have been re-investigated by high-resolution inelastic neutron scattering. We show that the changes go beyond a simple change in lifetime and frequency when the phonon frequency is close to the superconducting energy gap $2\Delta$. The observed lineshapes in elemental niobium are qualitatively similar to those found previously in borocarbide superconductors and agree very well with those predicted by the theory of Allen et al. [PRB \textbf{56}, 5552 (1997)]. Our results indicate that the peculiar phonon lineshapes in the superconducting state predicted by the theory of Allen et al. are a general phenomenon and not restricted to a particular class of compounds.
\end{abstract}

%\vspace{-0.8cm}
% body of paper here
% This is for two columns

  \maketitle

  \vskip2pc

\section{Introduction}
\label{intro}

Superconductivity-induced changes of the frequency and the linewidth of phonons with strong coupling to the electronic system have been observed in conventional superconductors a long time ago\cite{Axe73,Shapiro75}. Similar effects were later found in high-$T_c$ cuprate superconductors as well\cite{Krantz88,Friedl90,Pyka93}. When the phonon under investigation has a frequency smaller than the superconducting energy gap, there is a sudden drop in frequency as well as in linewidth on cooling through the superconducting transition temperature $T_c$. This effect can be understood on the basis that the coupling of such a phonon to the electrons vanishes in the superconducting state because phonons can no longer excite electron-hole quasi-particle pairs as soon as their energy is smaller than $2\Delta(T)$. For phonons with frequencies above $2\Delta$ an increase in linewidth was reported for $T<T_c$ (refs.\ \onlinecite{Axe73,Shapiro75}), which is due to the pile-up of electronic states just above $2\Delta$, i.e.\ the region in the electronic spectrum where the phonon couples to.

The early publications on superconductivity-induced changes of
phonon lineshapes concluded that these changes can be entirely
described by changes of phonon frequencies and phonon lifetimes.
In other words, it appeared that the intrinsic phonon lineshape
remains Lorentzian below $T_c$. However, marked deviations of the
phonon lineshape from a Lorentzian were later observed in
borocarbide superconductors, in particular in YNi$_{2}$B$_{2}$C
(ref.\ \onlinecite{Kawano96}) and LuNi$_{2}$B$_{2}$C (ref.\
\onlinecite{Stassis97}). For a particular acoustic phonon, the
lineshape below $T_c$ did so little resemble a Lorentzian that
the authors in Ref.\ \onlinecite{Kawano96} thought to have
observed a new type of excitations. These results motivated
theoretical work aiming at explaining the strange lineshapes
observed in experiment. In one of the theoretical papers, i.e.\
that of Kee and Varma\cite{Kee97}, anomalous lineshapes were
linked to a pole in the electronic polarizability in the
superconducting state that appears near extremum vectors of the
Fermi surface. This means that only phonons with wave vectors
corresponding to a nesting vector of the Fermi surface are
expected to show anomalous lineshapes in the superconducting
state.  The other theory put forward at that time, i.e.\ that by
Allen and co-workers\cite{Allen97}, predicted lineshapes as
observed in experiment quite generally, without any need for a
strong nesting.

The question which theory is more appropriate remained undecided for several years. Recently, we were able to show by studying phonons in YNi$_{2}$B$_{2}$C (ref.\ \onlinecite{Weber08}) with various momentum transfers that the theory of Allen et al.\cite{Allen97} gives a very good account of the experimental observations for all cases, which also means that nesting is not a prerequisite for the appearance of anomalous lineshapes in the superconducting state. Rather, the only relevant parameters are the ratios of the phonon frequency over the superconducting energy gap and of the linewidth in the normal state over the phonon frequency. Thus, similar effects as observed in borocarbide superconductors should show up in other superconductors as well, and one might ask why they have not been seen so far. When comparing the phonon properties of various superconductors with those of YNi$_{2}$B$_{2}$C we found that the latter compound is particularly favorable for such a study. In most other compounds, the predicted effects are so small that they are easily overlooked. That is to say, the effects are nearly or completely washed out by the resolution of available neutron spectrometers.

When searching for a superconductor other than the borocarbides
to check the theory of Allen et al.\cite{Allen97} we came upon
elemental Niobium, which has a lower $T_c$ than YNi$_{2}$B$_{2}$C,
i.e.\ $9.2\,\rm{K}$ instead of $15.2\,\rm{K}$, but still a fairly
strong electron-phonon coupling. The relatively strong
electron-phonon coupling in Nb was the reason that it was chosen
for an early inelastic neutron scattering study of the effects of
superconductivity on phonon frequencies and phonon lineshapes,
i.e.\ that of Shapiro et al.\cite{Shapiro75}. The authors of this
study did not mention to have observed any non-lorentzian
lineshapes. It seems that the instrumental resolution in this
study was insufficient to make any such deviation eye-catching.
Very recently, superconductivity-induced changes of phonon
linewidths have been investigated by a new technique, i.e
resonant spin-echo neutron spectroscopy\cite{Aynajian08}. This
technique is very powerful for detecting small intrinsic
lineshapes, much smaller than can be detected by classical
triple-axis neutron spectrometers. On the other hand, evaluation
of the raw data relies heavily on the assumption that the
intrinsic lineshape is lorentzian in nature, and any deviation
from a Lorentzian gives rise only to second order effects.
Therefore, it is understandable that any deviations from a
Lorentzian went unnoticed in this study.

In this paper, we use again classical triple-axis neutron spectroscopy to investigate phonons in Nb above and below $T_c$. We show that with improved resolution non-lorentzian lineshapes as predicted by theory are indeed observable. We further show that the theory of Allen et al.\cite{Allen97} gives a very good account of superconductivity-induced changes also for phonons with energies well below and above $2\Delta$, i.e.\ for phonons where the lineshape is not anomalous in the superconducting state, which further adds credence to this theory.

\section{Experimental}
 \label{experimental}
The neutron scattering experiments were performed on the 1T triple-axis spectrometer at the ORPHEE reactor at LLB, Saclay. Pyrolytic graphite was used both as monochromator and analyser. They were horizontally and vertically focusing, but tight collimation before and after the monochromator as well as before and after the analyser made the horizontal curvature ineffective. The tight collimation (25-20-20-30) was  necessary to achieve a sufficiently good resolution in energy and momentum space. In principle, it would have been desirable not to make use of the vertical curvature as well in order to improve the momentum resolution in the direction vertical to the scattering plane. As will be discussed below, the relaxed vertical resolution led to asymmetric phonon lineshapes. However, tightening the vertical resolution would have led to exceedingly long counting times and therefore, we had to adapt our analysis to take the effects of the relaxed vertical resolution into account. Achieving a good resolution required moreover to choose a low final energy, i.e.\ \wert{E_f}{5.0}{meV}. A graphite filter in the scattered beam was used to suppress higher orders.

The sample was a single crystal of Niobium having the shape of a cylinder of $1.2\,\rm{cm}$ in diameter and $2\,\rm{cm}$ high with its long axis being parallel to the crystallographic $(110)$ direction. Measurements were carried out in the $100-011$ scattering plane. The crystal was mounted in a standard orange cryostat allowing measurements down to \wert{T}{2}{K}.

\section{Results and Analysis}
\label{results}
  \begin{figure}
   \includegraphics[width=0.9\linewidth]{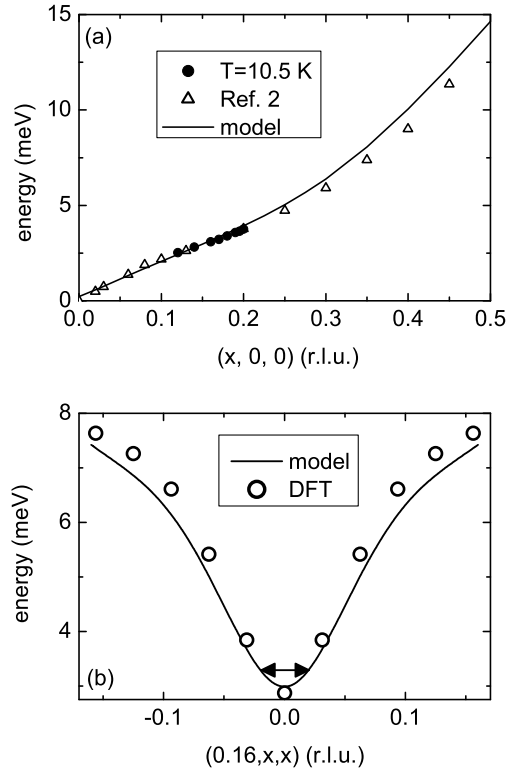}
   \caption{(a) Dispersion of the transverse acoustic phonon branch in the $(100)$ direction. The line was calculated from the force constant model published in Ref.\ \onlinecite{Reichardt01}. Filled symbols depict our results for \wert{T}{10.5}{K}, open symbols are taken from fig. 9 in Ref.\ \onlinecite{Shapiro75}. (b) Phonon dispersion in the off-symmetry direction $(0.16,x,x)$. The line was again calculated from the force constant model, whereas the open circles denote results of DFT calculations from Ref.\ \onlinecite{Pena-Seaman07}. They were shifted upwards by $0.9\,\rm{meV}$ in order to match the experimental value at $(0.16,0,0)$. The arrow denotes the momentum resolution in the direction vertical to the scattering plane.}
   \label{fig_1}
  \end{figure}

  \begin{figure}
   \includegraphics[width=0.9\linewidth]{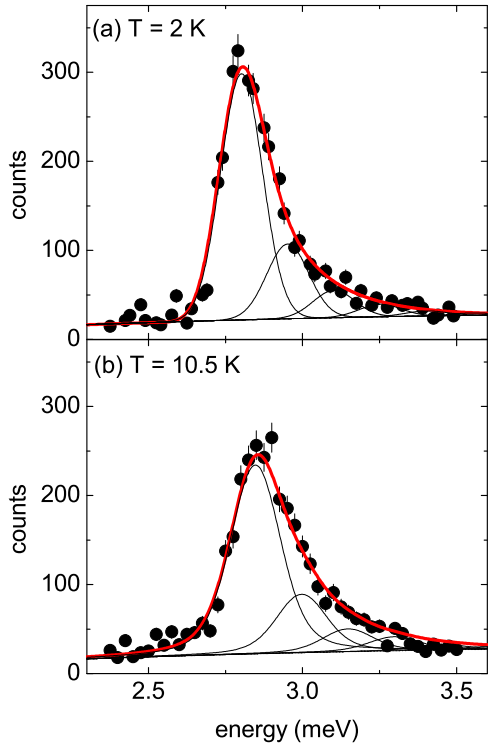}
   \caption{(color online) Energy  scans taken at \wert{\textbf{Q}}{(0.14, 1, 1)}{} at \wert{T}{2}{K} (a) and $10.5\,\rm{K}$ (b). The thick (red) line in (a) represents a simulation based on the lattice dynamical model of Ref.\ \onlinecite{Reichardt01} and the resolution in \textbf{Q} and $\omega$ of the spectrometer. The resulting spectrum was fitted by 5 Gaussians being $0.15\,\rm{meV}$ apart from each other (thin lines). (b) The thick (red) line was obtained by convoluting each Gaussian in (a) by the same Lorentzian and the same rigid energy shift to fit the normal state data. }
   \label{fig_2}
  \end{figure}

  \begin{figure}
   \includegraphics[width=0.9\linewidth]{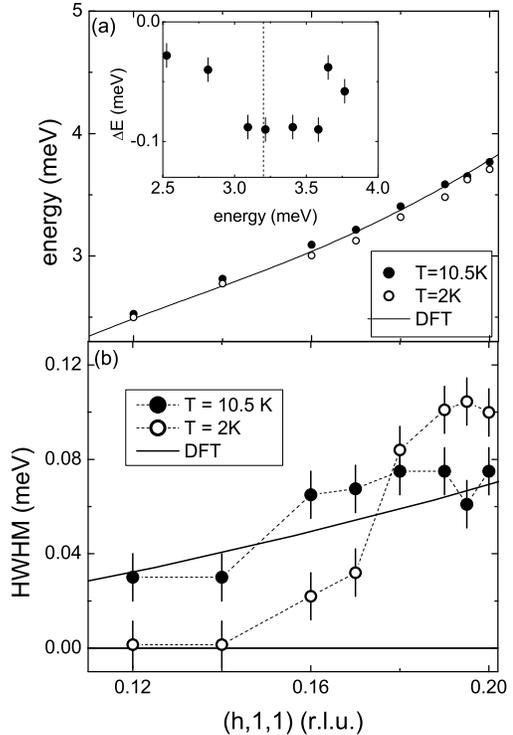}
   \caption{Frequency (a) and linewidth (half-width at half-maximum) (b) of transverse acoustic phonons in the $(100)$ direction observed above and far below $T_c$. The solid line in (a) represents DFT results\cite{Pena-Seaman07} multiplied by a factor $1.5$ to match the experimental data. The inset in panel (a) shows the observed energy shift $\Delta E$ on entering the superconducting state, while the dashed, vertical line indicates the superconducting energy gap \wert{2\Delta}{3.2}{meV}. The linewidth values were obtained by fitting the experimental spectra with a convolution of a Lorentzian and the resolution function of the spectrometer as indicated in Fig.~\ref{fig_2}(a). Note that deviations from a Lorentzian in the superconducting state as discussed in the text were not taken into account in these fits.}
   \label{fig_3}
  \end{figure}

  \begin{figure}
   \includegraphics[width=0.9\linewidth]{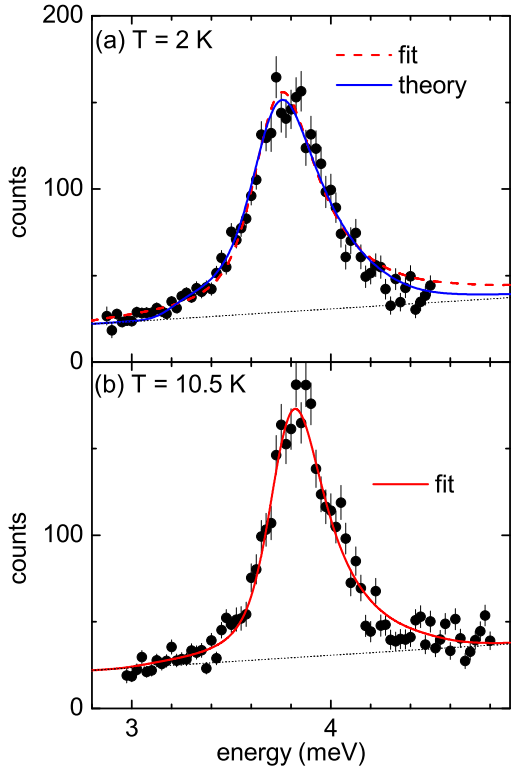}
   \caption{(color online) Energy scans of a phonon with a frequency well above the superconducting energy gap $2\Delta$ taken in the superconducting state (a) and in the normal state (b), respectively, at $\textbf{Q} = (0.2,1,1)$. The red lines depict fits by convoluting the resolution function of the spectrometer with a Lorentzian. The blue line was calculated from the theory of Allen et al.\cite{Allen97}. }
   \label{fig_4}
  \end{figure}

  \begin{figure}
   \includegraphics[width=0.92\linewidth]{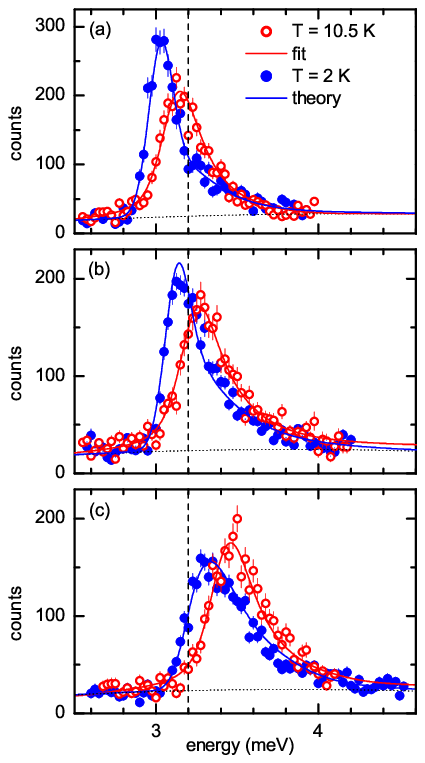}
   \caption{(color online) Energy scans taken at (a): $\textbf{Q} = (0.16,1,1)$, (b): $(0.17,1,1)$ and (c): $(0.18,1,1)$ at temperatures above (red) and below $T_c$ (blue). The fit curves represent a convolution of a Lorentzian with the resolution function. The theoretical curves were obtained using the theory of Allen et al.\cite{Allen97} and a value of the superconducting gap as indicated by the dashed vertical line. The experimental background is indicated by the thin horizontal lines. }
   \label{fig_5}
  \end{figure}

Our measurements were restricted to transverse acoustic phonons
propagating along $(100)$ for wave vectors between $q = 0.1$ and
$0.2$ (in units $2\pi/a$ with \wert{a}{3.27}{}\AA). From the
results published in Ref.\ \onlinecite{Shapiro75} we knew that
these phonons have a relatively large ratio of linewidth over
frequency and that the energies of these phonons would be
comparable to the superconducting energy gap. Hence these phonons
would be most favorable to search for superconductivity-induced
changes of the phonon lineshape. The dispersion of these phonons
is depicted in Fig.~\ref{fig_1}(a). These frequencies were
extracted from energy scans like that depicted in
Fig.~\ref{fig_2}. The asymmetry of the phonon line in
Fig.~\ref{fig_2}(a) is due to the finite momentum resolution in
conjunction with a steep increase of phonon frequencies when
going away from the high symmetry line (Fig.~\ref{fig_1}(b)). The
force constant model developed  by Reichardt and
Pintschovius\cite{Reichardt01} by fitting the parameters to data
published in the literature\cite{Powell68}  predicts a substantial frequency
increase within the region in momentum space sampled in our
experiment (Fig.~\ref{fig_1}(b)). This prediction is supported by
ab-initio calculations using density functional
theory\cite{Pena-Seaman07}. We used the phenomological model to
simulate our energy scans based on the formalism of Cooper and
Nathans\cite{Cooper67} for describing the four-dimensional
resolution function in $\textbf{q}$ and energy and found
quantitative agreement with experiment (Fig.~\ref{fig_2}(a)). We
note that the asymmetries related to resolution effects decrease
with increasing $q$ but are non-negligible up to the highest $q$
values studied here, i.e.\ up to $q = 0.2$.

In order to discriminate between asymmetries related to the finite momentum resolution and those related to superconductivity-induced distortions of the phonon lineshape, we adopted the following type of analysis: the phonon line predicted by the model was decomposed into five individual, equally spaced components having the linewidth expected for tight vertical resolution as illustrated in Fig.~\ref{fig_2}. Subsequently, each component was convoluted with the same Lorentzian to reproduce the lineshape in the normal state data. The Lorentzian linewidth obtained in this way (fig.~\ref{fig_3}(b)) was used as input for calculating the lineshape in the superconducting state within the framework of a theory published by Allen et al.\cite{Allen97}. The lineshape in the superconducting state predicted by this theory was calculated for each component separately and the sum was used for comparison with experiment. For the phonon depicted in Fig.~\ref{fig_2}, theory predicts  resolution-limited components somewhat down-shifted in energy. As will be discussed later, theory predicts complex lineshapes for other phonons, i.e.\ those with energies very close the superconducting energy gap.

For our analysis, we used the theory proposed by Allen et al.\cite{Allen97}, because we had found that it works quite well for the effects seen in borocarbide superconductors\cite{Weber08}. It is based on the full quantum mechanical treatment of electron-phonon coupling where vibrational and electronic excitations mix into hybrid modes. The theory contains three parameters, with two of them being entirely fixed by the phonon frequency and linewidth observed in the normal state. The third parameter is the superconducting energy gap $2\Delta$. The value of  $2\Delta$ was obtained by a comparison of the predicted to the observed phonon lineshape in the superconducting state. We note that the superconductivity-induced changes of the phonon spectra depend so sensitively on the gap value that the latter could be determined with a precision of about $1\,\%$.

In addition, we performed a simpler type of analysis to make contact with results of previous studies\cite{Shapiro75,Aynajian08}: in this case, the phonon spectra were fit with a single Lorentzian convoluted with the resolution function. From these fits, we obtained the superconductivity-induced changes of phonon frequencies and phonon linewidths depicted in Fig.~\ref{fig_3}.

The simple type of analysis just mentioned is fully adequate for
phonons with energies  well below or well above $2\Delta$: for
phonons with energies well below $2\Delta$, the phonon lineshape
becomes resolution limited below $T_c$ (Fig.~\ref{fig_2}(a)),
whereas for phonons with energies well above $2\Delta$ the
intrinsic lineshape remains essentially Lorentzian
(Fig.~\ref{fig_4}). A close look at the shape of the phonon line
depicted in Fig.~\ref{fig_4}(a) reveals that there are indeed
some discrepancies between the fit curve and the data, but they
are small. Therefore, it would be hard to argue on the basis of
such data that the lineshape in the superconducting state
deviates from a Lorentzian. Nevertheless, the curve calculated
from the theory of Allen et al.\cite{Allen97} describes the data
definitely better.

\begin{figure}
   \includegraphics[width=0.9\linewidth]{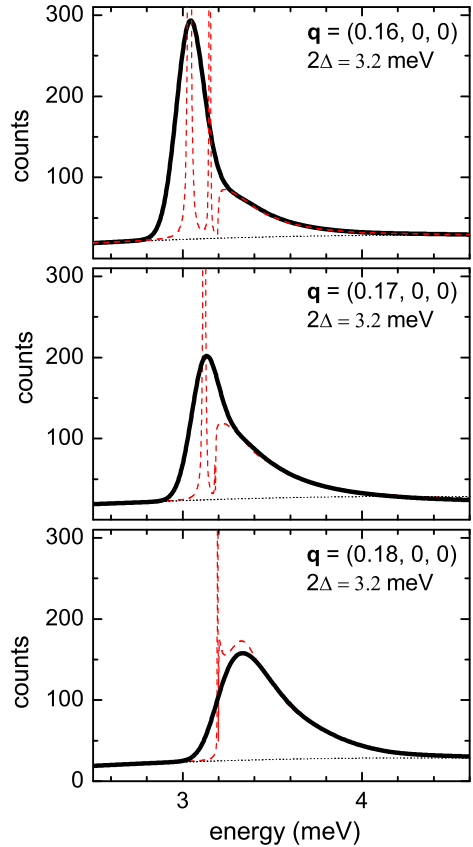}
   \caption{(color online) The dashed (red) lines and the full (black) ones depict the theoretical curves before and after convoluting them with the experimental energy resolution. The value of the superconducting energy gap used in the calculations is \wert{2\Delta}{3.2}{meV}. For the sake of comparison with fig.~\ref{fig_5}, the experimental background indicated by the horizontal dotted line has been added. The occurrence of two peaks in the top panel is an artifact related to the decomposition of the phonon line into a small number of individual components: a decomposition into a larger number of components would lead to densely spaced sharp peaks between $3\,\rm{meV} <  E < 3.2\,\rm{meV}$. However, the lineshape would barely change after convolution with the experimental energy resolution.}
   \label{fig_6}
  \end{figure}

We now turn to phonons with wavevectors $q = 0.16 - 0.18$ where the lineshape in the superconducting state deviates very clearly from a Lorentzian. The energies of these phonons are quite close to the value of the superconducting energy gap  \wert{2\Delta}{3.2}{meV}. We note that we tried larger and smaller values of  $2\Delta$ as well but found the best agreement between calculation and experiment for \wert{2\Delta}{(3.2 \pm 0.02)}{meV}, in agreement with the value reported in Ref.\ \onlinecite{Shapiro75}. The lineshapes of these phonons are depicted in Fig.~\ref{fig_5}. It is evident that the lineshape becomes markedly more asymmetric in the superconducting state, with a very steep intensity increase on the low energy side of the peak. Theory accounts both for the peculiar lineshape and the pronounced shift of the leading edge in a very satisfactory way. For a better understanding of the lineshapes predicted by theory, we show the theoretical curves before and after convolution with the experimental energy resolution in Fig.~\ref{fig_6}. Although the fine structure of the theoretical curves is washed out by the finite resolution, our data lend strong support to the basic features predicted by the theory of Allen et al.\cite{Allen97}.

\section{Discussion}
\label{discussion}

It is obvious from Fig.~\ref{fig_1} that the phonon frequencies obtained in our experiment match those reported in Ref.\ \onlinecite{Shapiro75} extremely well. Further, we note that the phonon linewidths evaluated in a conventional way (Fig.~\ref{fig_3}) agree with those reported in the two previous studies\cite{Shapiro75,Aynajian08} very well. A close look reveals that there is some discrepany between our values and those reported by Aynajian et al.\cite{Aynajian08} for the linewidths in the superconducting state around $q = 0.16$. Very probably, this discrepancy has a simple reason: that is to say, the measurements of Aynajian et al.\cite{Aynajian08} were done at a somewhat higher temperature than ours, i.e.\ $3.5\,\rm{K}$ instead of $1.8\,\rm{K}$ leading to somewhat larger linewidths.

The aim of the present study was to show that the conventional
analysis of the phonon lineshapes by fitting the data with a
Lorentzian is not fully adequate if the phonon energy is very
close to the superconducting energy gap $2\Delta$. Our data
obtained for wavevectors $q = 0.16 - 0.18$ clearly show that this
is the case, although the effects are less eye-catching than
those previously observed in borocarbide
superconductors\cite{Kawano96,Stassis97,Weber08}. It is not that
theory predicts much smaller effects for niobium than for the
borocarbides, but they occur on a much-reduced energy scale and
are therefore more washed out by the finite resolution of
available neutron spectrometers. Nevertheless, the very good
agreement between our observations and the curves calculated from
the theory of Allen et al.\cite{Allen97} confirm the predictions
of this theory: if the phonon energy in the normal state is
slightly above $2\Delta$, but its low energy tail extends below
$2\Delta$, the low energy tail will condense into a very sharp
peak right at $2\Delta$ on entering the superconducting state.
According to Allen et al.\cite{Allen97}, this sharp peak can be
regarded as a mixed vibrational/superelectronic excitation. If
the phonon energy in the normal state is right at $2\Delta$ or
even somewhat lower, the sharp peak is located somewhat below
$2\Delta$ and the high energy tail above $2\Delta$ has less
weight. If the phonon frequency in the normal state is
considerably smaller than $2\Delta$, the high energy tail becomes
vanishingly small, and the sharp peak corresponds to a normal
phonon with infinite lifetime. If the phonon energy in the normal
state is considerably larger than $2\Delta$, there is still an
intensity jump and a sharp peak right at $2\Delta$ but the jump
height is so small that it is very difficult if not impossible to
see it in experiment. Our data for $q = 0.20$ shown in
Fig.~\ref{fig_4} do support the presence of such an intensity
jump at $2\Delta$ but only marginally so, the jump height being
simply too small.

We emphasize that the theory of Allen et al.\cite{Allen97} does
not only explain the peculiar lineshapes observed for phonons
with energies very close to $2\Delta$ but also describes the
superconductivity-induced frequency shifts and linewidth changes
for phonons with smaller and larger frequencies in a quantitative
way. We note that the agreement between theory and experiment in
Niobium is better than previously found in YNi$_2$B$_2$C (Ref.\
\onlinecite{Weber08}). Possibly, the theory is better suited for
Nb because it was based on a weak-coupling BCS approach and the
electron-phonon coupling strength is indeed weaker in Nb than in
YNi$_2$B$_2$C. Another reason might be the fact that the
superconducting energy gap shows a marked anisotropy in
YNi$_2$B$_2$C (Ref.\ \onlinecite{Weber08}) but not in Nb, and
that the theory of Allen et al.\cite{Allen97} was developed for
the case of an isotropic gap.

\section{Conclusions}

We re-investigated the superconductivity-induced changes of the
phonon lineshape in elemental Niobium using high-resolution
inelastic neutron scattering. Our data confirm the results of
previous studies but also show unambiguously that for phonons
with energies very close to the superconducting energy gap, the
effects of superconductivity go beyond a simple change in
lifetime. Our results strongly support the view of Allen et al.\cite{Allen97} that, in contrast to a competing theory\cite{Kee97}, the peculiar phonon lineshapes observed in borocarbide superconductors are a general phenomenon and are not restricted to the case of a strong nesting effect. The effects found in Niobium are indeed quantitatively described by the theory
published by P. B. Allen et al.\cite{Allen97}. The agreement
between theory and experiment for Nb is better than for the
borocarbides. The theory does not only give a very good account
of the peculiar lineshapes of phonons with energies close to the
superconducting gap, but also of the changes in frequency and
linewidth for phonons with significantly smaller or larger
frequencies. The solid theoretical basis of the
superconductivity-induced changes of phonon lineshapes allows for
a precise determination of the superconducting energy gap in a
single phonon scan if the phonon energy is close to the gap
energy.

Work at Argonne National Laboratory was supported by the U.S. Department of Energy, Basic Energy Sciences-Materials Sciences, under Contract No. DE-AC02-06CH11357.

\end{document}